\documentclass[a4paper]{article}

\usepackage[english]{babel}
\usepackage[utf8x]{inputenc}
\usepackage{amsmath}
\usepackage{graphicx}
\usepackage[colorinlistoftodos]{todonotes}
\usepackage{hyperref}

\title{Dynamics of Coalescence-Induced \\ Jumping Water Droplets}
\author{Nenad Miljkovic,$^{1}$ \\Daniel J. Preston,$^{1}$ \\Ryan Enright,$^{2}$ \\Evelyn N. Wang$^{1}$ \\ \\
 $^{1}$Department of Mechanical Engineering,\\ Massachusetts Institute of Technology \\ Cambridge, MA 02139, USA \\ $^{2}$Thermal Management Research Group, \\ Bell Labs Ireland, Alcatel-Lucent, \\ Dublin 15, Ireland.}

\begin{document}
\maketitle

\begin{abstract}
This fluid dynamics video shows the different interaction mechanisms of coalescence-induced droplet jumping during condensation on a nanostructured superhydrophobic surface. High speed imaging was used to show jumping behavior on superhydrophobic copper oxide and carbon nanotube surfaces. Videos demonstrating multi-jumping droplets, jumping droplet return to the surface, and droplet-droplet electrostatic repulsions were analyzed. Experiments using external electric fields in conjunction with high speed imaging in a custom built experimental chamber were used to show that all coalescence-induced jumping droplets on superhydrophobic surfaces become positively charged upon leaving the surface, which is detailed in the video.
\end{abstract}

\section{Introduction}

With the broad interest and accelerating development of superhydrophobic surfaces for a variety of applications including self-cleaning, condensation heat transfer enhancement, and anti-icing, the need for more detailed insights on droplet interactions on these surfaces have emerged. Specifically, when two or more droplets coalesce, they can spontaneously jump away from a superhydrophobic surface independent of gravity due to the release of excess surface energy \cite{Kollera, Boreyko}. To date, researchers have focused on creating superhydrophobic surfaces showing rapid droplet removal and experimentally analyzing \cite{Miljkovic1} the merging and jumping behavior before and immediately after coalescence \cite{Enright}. However, aspects related to the droplet dynamics after departure from the surface remain to be investigated. Here, using high speed visualization, we show that jumping droplets 1) can undergo multiple jumps after departing the surface, 2) return to the surface against the force of gravity due to condensing vapor entrainment, and 3) gain a net positive charge that causes them to repel each other mid-flight \cite{Miljkovic2}. 

\section{Surface Fabrication}
To create the CuO nanostructures (Superhydrophobic CuO), commercially available oxygen-free Cu tubes were used (99.9{\%} purity) with outer diameters, {\it D}$_{OD}$ = 6.35 mm, inner diameters, {\it D}$_{ID}$ = 3.56 mm, and lengths, {\it L} = 131 mm, as the test samples for the experiments. Each Cu tube was cleaned in an ultrasonic bath with acetone for 10 minutes and rinsed with ethanol, isopropyl alcohol and de-ionized (DI) water. The tubes were then dipped into a 2.0 M hydrochloric acid solution for 10 minutes to remove the native oxide film on the surface, then triple-rinsed with DI water and dried with clean nitrogen gas. Nanostructured CuO films were formed by immersing the cleaned tubes (with ends capped) into a hot (96 ± 3$^\circ$C) alkaline solution composed of NaClO$_{2}$, NaOH, Na$_{3}$PO$_{4}$•12H$_{2}$O, and DI water (3.75 : 5 : 10 : 100 wt.{\%}). During the oxidation process, a thin (≈300 nm) Cu$_{2}$O layer was formed that then re-oxidized to form sharp, knife-like CuO oxide structures with heights of {\it h} $\approx$ 1 $\mu$m, solid fraction $\phi$ $\approx$ 0.023 and roughness factor {\it r} $\approx$ 10.

Carbon nanotubes (Superhydrophobic CNTs) were grown by chemical vapor deposition (CVD). Silicon growth substrates were prepared by sequentially depositing a 20 nm thick Al$_{2}$O$_{3}$ diffusion barrier and a 5 nm thick film of Fe catalyst layer using electron-beam deposition. Growth was performed in a 2.54 cm quartz furnace tube. Following a 15 min purge in a H$_{2}$/He atmosphere, the growth substrate was annealed by ramping the furnace temperature to 750$^\circ$C followed by a 3 minute anneal at temperature, while maintaining a flow of H$_{2}$ and He at 400 sccm and 100 sccm, respectively. CNT growth was then initiated by flowing C$_{2}$H$_{4}$ at 200 sccm. The flow of C$_{2}$H$_{4}$ was stopped after a period of 1 minute. The thermally-grown CNT had a typical outer diameter of {\it d} $\approx$ 7 nm. Due to the short growth time ($\approx$5 min.) the CNT did not form a well-aligned forest, but rather a tangled turf.

To functionalize the surfaces, a proprietary fluorinated polymer (P2i) was deposited using plasma enhanced vapor deposition. The process occurs under low pressure within a vacuum chamber at room temperature. The coating is introduced as a vapor and ionized. This process allows for the development of a highly conformal ($\approx$30 nm thick) polymer layer, which forms a covalent bond with the surface, making it extremely durable. Goniometric measurements (MCA-3, Kyowa Interface Science) of $\approx$100 nL droplets on a smooth P2i coated silicon wafer surface showed advancing and receding contact angles of $\theta$$_{a}$ = 124.3 ± 3.1$^\circ$ and $\theta$$_{r}$ = 112.6 ± 2.8$^\circ$, respectively.

\section{Experimental Setup and Conditions}
All experiments were carried out under saturated conditions in an environmental chamber. The droplet ejection process was captured using a single-camera set-up \cite{Miljkovic1}. The out-of-plane trajectory of the ejected droplets was captured using a high-speed camera (Phantom v7.1, Vision Research). The camera was mounted outside the environmental chamber and fitted with an extended macro lens assembly. The lens assembly consisted of a fully extended 5X optical zoom macro lens (MP-E 65mm, Canon), connected in series with 3 separate 68mm extension tubes (Auto Extension Tube Set DG, Kenko). The DG extension tubes have no optics. They are mounted in between the camera body and lens to create more distance between the lens and film plane. By moving the lens further away from the film or CCD sensor in the camera, the lens is forced to focus much closer than normal. The greater the length of the extension tube, the closer the lens can focus. Illumination was supplied by light emitting diodes installed inside the chamber and providing back lighting to the sample. The experiments were initiated by first evacuating the environmental chamber to medium-vacuum levels (=0.5 ± 0.025 Pa). Flat samples (jumping droplet videos (Superhydrophobic CuO and CNTs), and multi-jump videos) were mounted to a flattened copper tube connected to an external cooling loop and was maintained at a temperature of {\it T}$_{w}$ $\approx$ 26$^\circ$C ({\it p}$_{w}$ $\approx$ 3.33 kPa). The water vapor supply was vigorously boiled before the experiments to remove non-condensable gases. Water vapor was introduced into the environmental chamber {\it via} a metering valve set to maintain the chamber pressure. 

\subsection{Droplet Return Video}
Droplet return to the surface due to vapor flow entrainment was studied by observing steady state condensation on the nanostructured CuO tube captured with the high speed camera \cite{Miljkovic1}. The tube is oriented in the horizontal direction with cooling water flowing inside the tube at 5 L/min. The vapor pressure is ≈2.7 kPa. Droplet removal {\it via} coalescence-induced ejection occurs once droplets reach sizes large enough to begin coalescing. The video was captured at 90 fps and is played back at 30 fps. The field of view is 16.0 mm x 12.0 mm.

\subsection{Droplet-Droplet Repulsion Videos}
Droplet-droplet repulsion during steady state condensation on the nanostructured CuO tube captured with the high speed camera \cite{Miljkovic2}. The tube was oriented in the horizontal direction with the bottom surface seen on the top of the frame and cooling water flowing inside the tube at 5 L/min. The vapor pressure was ≈2.7 kPa. These videos were captured at 1000 fps and are played back at 5 fps. The fields of view for the two movies in succession are 2.8 x 2.9 mm, 3.3 x 2.6 mm, respectively.

\subsection{Charging Effects Videos}
To study the effect of droplet charging, a 350 $\mu$m diameter copper wire electrode was placed beneath the superhydrophobic surface \cite{Miljkovic2}. The electrode was connected to a 600 V DC power supply (N5752A, Agilent Technologies). The video shows a typical view from the side port of the tube-electrode setup after condensation initiated ($\Delta$V = 0 V). With an applied constant electrical bias ($\Delta$V), an electric field between the electrode and grounded tube was established, inducing droplet motion toward or away from the electrode. High speed videos show droplet motion in the presence of the electrode. When a negative bias was applied to the electrode ($\Delta$V = -15, -30 V), significant droplet-electrode attraction was observed. To eliminate the possibility of induced electrical effects, {\it i.e.}, droplet motion due to dielectrophoresis, we reversed the polarity of the electrode ($\Delta$V = +15, +30 V) and saw a significant droplet-electrode repulsion. The repulsion and attraction observed under positive and negative electrode bias, respectively, indicate that all of the droplets were positively charged after jumping from the surface.\\

The visualizations provide insight into complex droplet-vapor, droplet-surface, and droplet-droplet phenomena, which offer a realm of new possibilities.

\end{document}